\documentclass[apj]{emulateapj}
\usepackage{amsmath}
\usepackage{amsbsy}
\usepackage{apjfonts}
\usepackage{multirow}
\usepackage{array}

\newcommand{\DDir}{\relax{D\kern-.7em{/}}}

\newcommand{\haf}{\frac{1}{2}}
\newcommand{\inv}[1]{\frac{1}{#1}}

\newcommand{\ra}{\rightarrow}

\newcommand{\xra}{\xrightarrow}
\newcommand{\soo}{\Rightarrow}

\newcommand{\be}{\begin{equation}}
\newcommand{\ee}{\end{equation}}
\newcommand{\bea}{\begin{equation*}}
\newcommand{\eea}{\end{equation*}}

\newcommand{\ave}[1]{\left\langle #1\right\rangle}

\newcommand{\pr}{\partial}

\newcommand{\nin}{\relax{\in\kern-.8em{/}}}

\newcommand{\te}{\theta}
\newcommand{\al}{\alpha}
\newcommand{\bt}{\beta}

\newcommand{\De}{\Delta}

\newcommand{\Om}{\Omega}

\newcommand{\sig}{\sigma}

\newcommand{\vt}{\textrm{v}}

\newcommand{\cm}{\mbox{ cm}}

\newcommand{\se}{\mbox{ s}}

\newcommand{\erg}{\mbox{ erg}}

\newcommand{\km}{\mbox{ km}}

\newcommand{\keV}{\mbox{ keV}}

\newcommand{\gr}{\mbox{g}}

\newcommand{\hr}{\mbox{hr}}
\newcommand{\sref}{\S~\ref}

\newcommand{\Lobs}{L_{\rm obs}}

\newcommand{\tref}{t_{\rm ref}}
\newcommand{\tpeakj}{\tref}
\newcommand{\tobs}{t_{\rm obs}}

\newcommand{\bOm}{\boldsymbol\Om}

\newcommand{\mucom}{\mu_{\rm com}}
\newcommand{\Icom}{I_{\rm com}}
\newcommand{\Lcom}{L_{\rm com}}

\begin{document}
\title{Non-relativistic radiation mediated shock breakouts: \\II. Bolometric properties of SN shock breakout}
\author{Boaz Katz\altaffilmark{1,2}, Nir Sapir\altaffilmark{3}, and Eli Waxman\altaffilmark{3}}

\altaffiltext{1}{Institute for Advanced Study, Princeton, NJ 08540, USA}
\altaffiltext{2}{John Bahcall Fellow, Einstein Fellow}
\altaffiltext{3}{Dept. of Particle Physics \& Astrophysics, Weizmann Institute, Rehovot 76100, Israel}

\begin{abstract}
Exact bolometric light curves of supernova shock breakouts are derived based on the universal, non relativistic, planar breakout solutions \citep{Sapir11}, assuming spherical symmetry, constant Thomson scattering opacity, $\kappa$, and angular intensity corresponding to the steady state planar limit. These approximations are accurate for progenitors with a scale height much smaller than the radius. The light curves are insensitive to the density profile and are determined by the progenitor radius $R$, and the breakout velocity and density, $\vt_0$ and $\rho_0$ respectively, and $\kappa$. The total breakout energy, $E_{\rm BO}$, and the maximal ejecta velocity, $\vt_{\max}$, are shown to be $E_{\rm BO}=8.0\pi R^2\kappa^{-1}c\vt_{0}$ and $\vt_{\max}=2.0\vt_{0}$ respectively, to an accuracy of about $10\%$.  The calculated light curves are valid up to the time of transition to spherical expansion,  $t_{\rm sph}\approx R/4\vt_0$. Approximate analytic expressions for the light curves are provided for breakouts in which the shock crossing time at breakout, $t_0=c/\kappa\rho_0\vt_0^2$, is $\ll R/c$ (valid for $R<10^{14}$~cm).  Modifications of the flux angular intensity distribution and differences in shock arrival times to the surface, $\Delta t_{\rm asym}$, due to moderately asymmetric explosions, affect the early light curve but do not affect $\vt_{\max}$ and $E_{\rm BO}$. For $4\vt_0\ll c$, valid for large (RSG) progenitors, $L\propto t^{-4/3}$ at $\max(\Delta t_{\rm asym},R/c)< t<t_{\rm sph}$ and $R$ may be accurately estimated from $R\approx 2\times 10^{13} (L/10^{43}\erg\sec^{-1})^{2/5}(t/1\hr)^{8/15}$.
\end{abstract}

\section{Introduction}
\label{sec:intro}

During a core collapse supernova (SN) explosion, a strong radiation mediated shock (RMS) traverses the exploding stars' mantle/envelope. Once the shock reaches the surface of the star, a burst of high energy radiation (UV to $\gamma$-rays) is expected to be emitted \citep[][]{Colgate74b,Falk78,Klein78,Ensman92,Matzner99,Blinnikov00,Katz09,Piro10,Nakar10}.

The observed properties of this breakout were derived in earlier analyses using either analytic order of magnitude estimates \citep[e.g.][]{Matzner99,Katz09,Piro10,Nakar10} or numerical calculations for particular progenitors \citep[e.g.][]{Ensman92,Blinnikov00,Utrobin07,Tominaga09,Tolstov10,Dessart10,Kasen11}. Here we provide an accurate description of the time dependent radiation emission following shock breakout for a general progenitor without an optically thick wind.  

In the first paper of this series \citep[][hereafter Paper I]{Sapir11} we have solved the problem of a non-steady planar RMS breaking out from a surface with a power-law density profile, $\rho\propto x^n$, in the approximation of diffusion with constant opacity \cite[a similar solution for exponential atmospheres, valid only for the early part of the planar breakout, was given in][]{Lasher79}. In this paper we use the results of the planar calculation to derive the observed bolometric properties of SN shock breakout bursts, taking into account limb darkening. In a third paper \citep[][]{Sapir11b} we calculate the temperature profiles and spectral properties of the burst assuming local Compton equilibrium and photon generation by Bremsstrahlung (see \S~5 of paper I).

The derivation of exact light curves for a general progenitor is possible thanks to the universality of the planar breakout solutions (Paper I). Radiation escapes the shock front, producing the observed breakout burst, when the optical depth $\tau$ of the plasma lying ahead of it is equal to the optical depth of the shock transition layer, $\tau_{\rm sh}=c/\vt_{\rm sh}$ \citep{Weaver76}. We denote the shock velocity and (pre-shock) density at this point by $\vt_0$ and $\rho_0$ respectively (see \S~\ref{sec:Planar} for exact definitions). Measuring length, time and mass in units of $x_0=c/\kappa\rho_0\vt_0$, $t_0=x_0/\vt_0$ and $m_0/x_0^2=c/\kappa\vt_0$ respectively, the RMS breakout solution is universal and depends only on the density power law index $n$. It was numerically found that the luminosity depends weakly on the density structure. In fact, throughout most of the emission, the luminosity changes by less than $25\%$ for $n$ in the range $1-10$. The fact that the planar luminosity curve changes so little for a wide range of density power law indexes suggests that the results are insensitive to the decreasing density structure, and are applicable to profiles which are not power laws.

The results presented in this paper, based on the non-relativistic planar breakout solution, have corrections of order $\vt_0/c$. Estimating these and higher order corrections require a relativistic calculation of a planar shock breakout, and are beyond the scope of this paper. We do estimate one obvious first order correction resulting from the transformation of the comoving calculated flux to the observer frame in \sref{sec:FirstOrder}. An additional complication occurs at shock velocities $\vt/c\gtrsim0.3$, where the post shock temperatures reach $50\keV$ \citep{Weaver76,Katz09} and electron-positron pairs are created and increase the opacity of the material. Our discussion is applicable to smaller velocities/lower temperatures. Order of magnitude estimates for relativistic shock breakouts can be obtained by using the steady state solution of a relativistic shock \citep{Budnik10,Katz09}. 

This paper is organized as follows. The planar non-relativistic RMS breakout solution is shortly described in \sref{sec:Planar}. The application of the planar solution to a SN breakout is discussed in \sref{sec:Planar_SNe}: The planar breakout parameters $\vt_0$ and $\rho_0$ are expressed in terms of progenitor properties and explosion energy, and complications due to spherical expansion, asymmetry and relativistic corrections are discussed. Numerically calibrated analytic expressions for the total breakout energy, $E_{\rm BO}$, the asymptotic velocity of the fastest moving ejecta, $\vt_{\max}$, and the luminosity at late times, $t>\max(\Delta t_{\rm asym},R/c)$, are given in \sref{sec:EnergyVelocityLuminosity}. These are all insensitive to small deviations from spherical symmetry and from the constant flux angular intensity distribution. In \sref{sec:lightcurve}, the observed bolometric light curve during breakout is calculated, assuming that the shock arrives at the surface simultaneously and taking into account light travel time effects. Approximate analytic expressions for the light curve are derived. Some of the first order corrections in $\vt/c$ to the burst properties are analyzed in \sref{sec:FirstOrder}. Our results are compared to those of previous studies in \sref{sec:Comparison}. The main results and conclusions are summarized in \sref{sec:Summary}. Appendices \sref{sec:gs}-\sref{sec:Chandra} provide details omitted from the manuscript.

\section{Planar non-relativistic RMS breakout}
\label{sec:Planar}

The problem of a non relativistic RMS breaking out from the surface of a planar decreasing density profile was solved in Paper I \citep[see also][]{Lasher79}. The analysis is preformed by neglecting the thermal energy of the matter and by approximating the radiation energy transport by diffusion with constant opacity $\kappa$.

The initial density profile is assumed to be a power-law, $\rho\propto x^n$, where $x$ is the distance from the surface. The initial density and the asymptotic shock velocity at large optical depth are parameterized by
\begin{equation}\label{eq:InitialProfileRho}
\rho(\tau)=\rho_0 (\vt_0\tau/c)^{n/(n+1)}
\end{equation}
and
\begin{equation}\label{eq:InitialProfileV}
\vt_s(\tau)\xra[\tau\ra\infty]{}\vt_0 (\vt_0\tau/c)^{-\bt_n n/(n+1)}
\end{equation}
respectively, where $\tau=\kappa\int \rho dx$ is measured with respect to the surface and where $\bt_n$ is a function of $n$, which can be obtained numerically by solving the pure hydrodynamic shock evolution \citep{Sakurai60}. The values of $\rho_0$ and $\vt_0$ are the density and velocity at the point $\tau=c/\vt_{\rm sh}$  that would have been obtained in a pure hydrodynamic shock propagation.

The limits $n\ra0$ and $n\ra\infty$ are well defined and correspond to a homogeneous density distribution and an exponential profile \citep{Hayes68},
\begin{equation}
\rho\xra[n\ra\infty]{}\rho_0\exp(\kappa\rho_0\De x),\quad \De x=x(\rho)-x(\rho_0)
\end{equation}
respectively. The value of $\bt_n$  in Eq. \eqref{eq:InitialProfileV} decreases monotonically from $0.207$ to $0.176$ for $0<n<\infty$.

As mentioned in \S~\ref{sec:intro}, measuring length, time and mass in units of $x_0=c/\kappa\rho_0\vt_0$, $t_0=x_0/\vt_0$ and $m_0/x_0^2=c/\kappa\vt_0$ respectively, the RMS breakout solution is universal, i.e. depends on the dimensionless parameter $n$ alone. In particular, the planar emitted energy flux, i.e. the luminosity per unit area, can be expressed as
\begin{equation}\label{eq:L_planar}
    \mathcal{L}(t)=\mathcal{L}_0 \mathcal{\tilde L}\left(\frac{t-\tpeakj}{t_0},n\right),
\end{equation}
where $\mathcal{L}_0=\rho_0\vt_0^3$ and $\mathcal{\tilde L}\equiv \mathcal{L}/\mathcal{L}_0$ is given in table 3 of Paper I for $n$ in the range $1-10$. The shock crossing time at breakout is defined as
\begin{equation}\label{eq:tNorm}
t_0=\frac{c}{\kappa \rho_0 \vt_0^2}.
\end{equation}
The reference time $\tpeakj$ corresponds to $t_{\rm peak}$ in Paper I, the time at which the luminosity per unit area $\mathcal{L}$ peaks in the planar solution. Since the observed SN light curves do not peak at the same time (due to light travel time effects) we have replaced $t_{\rm peak}$ with  $\tpeakj$ to avoid confusion.

As can be seen in table 3 of Paper I, the luminosity depends weakly on the density structure. In fact, throughout most of the emission, the luminosity changes by less than $25\%$ for $n$ in the range $1-10$. The fact that the planar luminosity curve changes so little for a wide range of density power law indexes suggests that the result is insensitive to the decreasing density structure, and likely represents also profiles which deviate from a power law structure.

For convenience, we include a surface area of $4\pi R^2$ in the expressions below. The luminosity, $L(t)=4\pi R^2 \mathcal{L}(t)$, may be written as
\begin{equation}\label{eq:LNorm}
L(t)=L_0 \mathcal{\tilde L}\left(\frac{t-\tpeakj}{t_0},n\right),
\end{equation}
where $L_0$ is the breakout luminosity defined by
\be\label{eq:L_0def} L_0\equiv4\pi R^2 \rho_0\vt_0^3.\ee
At late times, $t\gg t_0$, the luminosity follows $L(t)\propto t^{-4/3}$ \citep{Piro10,Nakar10}\footnote{for a constant density profile, $n=0$, the decline is steeper, $L(t)\propto t^{-9/8}$ \citep{Sapir11}}. In Paper I it was found that the exact solutions for the luminosity and its integral, $E(t)=\int^t L(t') dt'$, are well approximated by
\begin{align}\label{eq:ELApprox}
&L(t)= L_{\infty}\left(\frac{t-\tpeakj}{t_0}\right)^{-4/3},\cr
&E(t)= E_{\infty}\left[1-\left(\frac{t-\tpeakj}{a_t t_0}\right)^{-1/3}\right],\cr
\end{align}
where
\begin{align}\label{eq:Einfty}
&E_{\infty}=2.0\times 4\pi R^2\frac{\vt_0c}{\kappa},\cr
&L_{\infty}=0.33\times 4\pi R^2\rho_0\vt_0^3,\cr
&a_t=(3L_{\infty}t_0/E_{\infty})^3=0.125.
\end{align}
Eqs. \eqref{eq:ELApprox} and \eqref{eq:Einfty} describe the emitted flux to an accuracy of better than  $30\%$ ($10\%$) in $L(t)$ ($E(t)$)  for $1<n<10$ and $1<(t-\tpeakj)/t_0<100$. Individual fits to different values of $n$ allow higher accuracy.
Note that $E_{\infty}$ is the total energy emitted in the planar approximation.

Finally, in the non relativistic approximation in planar geometry an exact relation exists between the velocity of the outermost mass element and the emitted luminosity \citep{Lasher79, Sapir11},
\begin{equation}\label{eq:v_to_Egen}
\vt(t)=\frac{\kappa}{c}\int_{-\infty}^t\mathcal{L}(t')dt'=\frac{\kappa E(t)}{4\pi R^2c}.
\end{equation}
In particular, the asymptotic value of the velocity of the surface is
\begin{equation}\label{eq. Assymptotic Velocity}
\vt_{\infty}=\frac{\kappa E_{\infty}}{4\pi R^2c}= 2.0\vt_0.
\end{equation}
Equation \eqref{eq:v_to_Egen} simply states that photons that hit a given particle transfer all their momentum to the particle on average. It holds for any elastic scattering which has forward/backward symmetry, regardless of whether the diffusion approximation is valid or not.

\section{Application of the planar solution to a SN breakout}
\label{sec:Planar_SNe}

In this section the application of the solution of shock breakout in the planar approximation to supernova shock breakouts is discussed. In \sref{sec:BreakoutParameters} the relation between the progenitor parameters and the breakout parameters is briefly reviewed. The effects that need to be taken into account in calculating the light curves using the planar solutions are described in \sref{sec:PlanarApp}. The limitations of the planar approximation are discussed in \S~\ref{sec:limitations}.

\subsection{Breakout parameters}
\label{sec:BreakoutParameters}

The relation between the parameters at breakout and the physical parameters of the SN explosion were given in numerous publications \citep[e.g.][]{Matzner99,Katz09,Nakar10}. In particular, a complete set of such relations is given in Appendix A of \citet[][]{Nakar10}. For convenience, we reproduce the relation for $\vt_0$ and $\rho_0$ below. As explained in \S~\ref{sec:Planar}, these quantities completely define the planar problem.

We use the approximate relation for the evolution of the shock velocity throughout the star \citep[Eq. 19 in][]{Matzner99} to set \begin{equation}\label{eq:v0_norm}
    \vt_{0}\approx1.0\vt_*[\bar\rho/\rho_0]^{0.19},
\end{equation}
with
\begin{equation}\label{eq:v*}
    \vt_*=(E_{\rm in}/M_{\rm ej})^{1/2}, \quad \bar \rho=M_{\rm ej}/(4\pi R^3/3).
\end{equation}
Here, $M_{\rm ej}$ is the mass of the ejecta and $E_{\rm in}$ its energy (note that $\bar\rho$ is different from $\rho_*=M_{\rm ej}R^{-3}$ by a factor of $4\pi/3$). We further use the density parametrization $\rho(x)=f_{\rho}\bar \rho (x/R)^{n}$, where $f_\rho$ is a dimensionless parameter of order  unity \citep[see][appendix A for detailed estimates, note that $\rho_1/\rho_*$ defined by the authors is related to $f_{\rho}$ defined here by $f_{\rho}=4\pi/3(\rho_1/\rho_*)$]{Calzavara04}.  Solving for $\tau=\bt_0^{-1}$, the following relations are obtained for $n=3$ (appropriate for a blue supergiant (BSG)) and for $n=3/2$ (appropriate for a red supergiant (RSG)):

\begin{align}\label{eq:MVParam_v}
\vt_0/\vt_*&= 13 M_{10}^{0.16}\vt_{*,8.5}^{0.16}R_{12}^{-0.32}\kappa_{0.4}^{0.16} f_{\rho}^{-0.05}~~(BSG)\cr
&= 4.5 M_{10}^{0.13}\vt_{*,8.5}^{0.13}R_{13}^{-0.26}\kappa_{0.4}^{0.13} f_{\rho}^{-0.09}~~(RSG),
\end{align}
\begin{align}\label{eq:MVParam_rho}
\rho_0&= 7\times 10^{-9} M_{10}^{0.13}\vt_{*,8.5}^{-0.87}R_{12}^{-1.26}\kappa_{0.4}^{-0.87} f_{\rho}^{0.29} \gr \cm^{-3}~~(BSG)\cr
&= 2 \times 10^{-9} M_{10}^{0.32}\vt_{*,8.5}^{-0.68}R_{13}^{-1.64}\kappa_{0.4}^{-0.68} f_{\rho}^{0.45} \gr \cm^{-3}~~(RSG),
\end{align}
where $M_{\rm ej}=10 M_{10} M_{\odot}$, $R=10^{12}R_{12}\cm=10^{13}R_{13}\cm$, and $\vt_*=3,000\vt_{*,8.5}\km\se^{-1}$.

\subsection{Applicability of the planar solution to a SN breakout}\label{sec:PlanarApp}
Several effects must be taken into account when using the planar solution to describe the observed breakout burst.

\paragraph{Light travel time smearing}  A distant observer sees the breakout emission coming from different locations on the surface at different times due to the finite light travel time. This can be accounted for by appropriately "smearing" the instantaneous luminosity on a time scale $t_{\rm smear}\sim R/c$,  as described in \sref{sec:lightcurve}.

\paragraph{Relativistic corrections} At high velocities, $\bt_0\equiv\vt_0/c\gtrsim0.1$, relativistic effects may introduce corrections of order tens of percents to the observed luminosity. One aspect which is easily accounted for are first order corrections to the relation between the calculated comoving fluxes and the observed fluxes. These corrections can be accounted for by Lorentz transforming the quantities in the comoving frame to the laboratory frame and finding the retarded time $t_{\rm ret}$ of the emission of photons arriving at the observer at time $t_{\rm obs}$. This is discussed in \sref{sec:FirstOrder}.

\subsection{Limitations}
\label{sec:limitations}

The applicability of the planar solution is limited by the following complications.
\paragraph{Spherical expansion}  As the outer mass elements expand, their optical depth decreases like $\tau\propto r^{-2}$, where $r(m,t)$ is the radius to which the mass element moved. The planar approximation breaks once the optical depth of the outermost elements drops significantly. Given that the outermost mass elements move with a velocity $\vt\approx 2\vt_0$, the optical depth of the outermost elements drops by a factor of $\sim2$ at
\begin{equation}\label{eq:t_sph}
t_{\rm sph}\sim \frac{\sqrt{2}-1}{2}\frac{R}{\vt_0}\sim R/(4\vt_0).
\end{equation}
The use of the planar solution is limited to times $t\ll t_{\rm sph}$.

For the planar solution to be applicable, is is required that $t_{\rm sph}\gg t_0$ (this is equivalent to $R\gg x_0$). Using Eqs. \eqref{eq:MVParam_v}-\eqref{eq:t_sph} we have
\begin{align}\label{eq:t_0t_sph}
t_0/t_{\rm sph}&\sim0.01 M_{10}^{-0.29}\vt_{*,8.5}^{-0.29}R_{12}^{0.58}\kappa_{0.4}^{-0.29} f_{\rho}^{-0.24}&(BSG)\cr
&\sim0.01 M_{10}^{-0.45}\vt_{*,8.5}^{-0.45}R_{13}^{0.9}\kappa_{0.4}^{-0.45} f_{\rho}^{-0.37}&(RSG)
\end{align}
implying that $t_0\ll t_{\rm sph}$ for practically all progenitors.

We emphasize  that our results are not applicable for progenitors with optically thick winds.

\paragraph{Non spherically symmetric explosions}  It is likely that SN explosions are not spherically symmetric. The use of the planar solution for asymmetric explosions may be limited due to several effects:
\begin{enumerate}
\item The break out velocities may be different at different locations on the surface;
\item The shock may reach the surface at oblique angles;
\item The shock arrival time to the surface may depend on location.
\end{enumerate}
The treatment of the first two effects is beyond the scope of this paper.
If the shock arrives to the surface at a large angle, the planar symmetry does not hold and our solution is not applicable.  As the shock propagates through the star the anisotropy is expected to be smoothed, and it is reasonable to expect that there is a wide range of parameters for which the velocities are approximately uniform and the obliqueness is small. We note that even if the velocities are different, our solution can be applied locally, with the local value of the shock velocity, at any location where the shock arrives at small obliqueness.

The third problem can, in principle, be considered within the context of the planar solution. It amounts to an appropriate smearing of the instantaneous luminosity over a time scale $\Delta t_{\rm asym}$ spanning the arrival of the shock to different locations. Section \sref{sec:EnergyVelocityLuminosity} discusses properties that are not affected by this complication or by light travel smearing.

\section{Breakout energy, maximal ejecta velocity and asymptotic luminosity}\label{sec:EnergyVelocityLuminosity}
In this section, robust approximate expressions are given for the total breakout energy, the velocity of the fastest moving elements, and the luminosity at late times, $\max(\Delta t_{\rm asym},R/c)<t<R/(4\vt_0)$. These expressions are insensitive to light travel time averaging and to deviations from instantaneous arrival of the shock to the surface at all location.

\subsection{Breakout energy}
The breakout energy, defined as the total emitted energy up to the time of transition to spherical expansion, is given by (using Eq. \eqref{eq:ELApprox})
\begin{equation}\label{eq:EBreakout}
E_{\rm BO}=E_{\infty}\left[1-\left(\frac{t_{\rm sph}}{a_t t_0}\right)^{-1/3}\right].
\end{equation}
Using equation \eqref{eq:t_0t_sph} and $a_t\sim0.1$, the second term in the parenthesis is found to be of order $0.1$ and to a good approximation $E_{\rm BO}\approx E_{\infty}$.

The breakout energy is not sensitive to the precise value taken for the time of transition to spherical
expansion, as we now illustrate. Beyond the transition to spherical expansion, the luminosity will level off approaching an asymptotic power law
\begin{equation}
L(t)\propto L_{\rm sph}(t/t_{\rm sph})^{-\alpha_{\rm sph}},
\end{equation}
where $L_{\rm sph} =L(t_{\rm sph} )$ is the luminosity at $t_{\rm sph}$
and $\alpha_{\rm sph}=0.34(0.17)$ for $n=3 (3/2)$ \citep{Chevalier92,Rabinak10,Nakar10}. It is useful to estimate the time at which the contribution of the later spherical phase emission, $\Delta E$, is comparable to $E_{\rm BO}$. The time it takes to accumulate an energy $E_{\rm BO}$ in the spherical phase is roughly given by
\begin{align}
\Delta t_{\rm BO}&= t_{\rm sph}\left[(1-\al_{\rm sph})E_{\rm BO}/(L_{\rm sph}t_{\rm sph})\right]^{1/(1-\al_{\rm sph})}\cr
&= 3t_{\rm sph}\left(\frac{t_{\rm sph}}{a_t t_0}\right)^{1/(3-3\al_{\rm sph})}.
\end{align}
Using Eqs.~\eqref{eq:t_sph},~\eqref{eq:t_0t_sph} and  \eqref{eq:MVParam_v} we find
\begin{align}\label{eq:DtBr}
\Delta t_{\rm BO}\vt_*/R&= 1.8 M_{10}^{-0.01}\vt_{*,8.5}^{-0.01}R_{12}^{0.03}\kappa_{0.4}^{-0.01} f_{\rho}^{0.17}~~(BSG)\cr
&= 2.4 M_{10}^{0.05}\vt_{*,8.5}^{0.05}R_{13}^{-0.1}\kappa_{0.4}^{0.05} f_{\rho}^{0.23}~~(RSG),
\end{align}
implying that for all progenitors
\begin{equation}\label{eq:DtBrNum}
\Delta t_{\rm BO}\sim 2 R/\vt_{*}= 18 R_{13}\vt_{*,8.5}^{-1}\hr\gg t_{\rm sph}.
\end{equation}
At earlier times, the accumulated energy emitted in the spherical phase, $\De E$, relative to the breakout energy, is roughly given by
\begin{equation}
\frac{\De E}{E_{\rm BO}}= \left(\frac{t}{\Delta t_{\rm BO}}\right)^{1-\al_{\rm sph}}.
\end{equation}
The difference in shock arrival time to the surface due to asymmetry is always much shorter than $R/\vt_*$, implying that the breakout energy can be accurately integrated, even if the time difference in arrival times is significant.

\subsection{Maximal velocity}
The velocity of the fastest moving ejecta can be obtained from the planar
relation Eq. \eqref{eq:v_to_Egen}. This velocity can be probed by other observations, including the spectrum of the breakout and radio observations of the collisionless shock propagating through the circumstellar medium \citep[e.g.][]{Chevalier06,Waxman07}. As long as the planar approximation is valid, the velocity of the surface is proportional to the emitted energy and is given by Eq. \eqref{eq:v_to_Egen}. Once the radius changes considerably, the acceleration declines sharply, $\pr_t \vt\propto LR^{-2}$,  and the velocity does not increase significantly any more \citep[see also][]{Matzner99}.
The resulting velocity of the fastest part of the ejecta is thus related to the breakout energy
\begin{equation}\label{eq:vmaxEBr}
\vt_{\max}= \frac{\kappa}{4\pi R^2 c}E_{\rm BO},
\end{equation}
and is given by
\begin{equation}\label{eq:vmax}
\vt_{\max}= 2.0\vt_0\left[1-\left(\frac{t_{\rm sph}}{a_t t_0}\right)^{-1/3}\right].
\end{equation}

A rough estimate of the additional acceleration beyond the transition to spherical expansion can be obtained by approximating $L(t>t_{\rm sph})=\rm const=\rm L_{\rm sph}$ and $R\propto t$. In this limit, the acceleration drops like $t^{-2}$ beyond $t_{\rm sph}$ and the total additional velocity $\De \vt_{\max}$ is roughly given by [using Eq. \eqref{eq:EBreakout}]
\begin{equation}
\frac{\De \vt_{\max}}{\vt_{\max}}\sim \frac{t_{\rm sph}L_{\rm sph}}{E_{\rm BO}}\sim \inv{3}\left(\frac{t_{\rm sph}}{a_t t_0}\right)^{-1/3},
\end{equation}
implying a correction of a few percent for all progenitors considered.

\subsection{Asymptotic luminosity}\label{sec:AssymVel}
In the regime $t_{\rm smear}\ll t\ll t_{\rm sph}$, where $t_{\rm smear}=\max(\Delta t_{\rm asym},R/c)$, the luminosity is given by equation \eqref{eq:ELApprox} and can be expressed as
\begin{align}\label{eq:Lasymp}
L(t)&= L_{\infty}(t/t_0)^{-4/3}\cr
&= \frac4{3.0}\pi R^2\kappa^{-4/3}\left(\frac{\vt_0}{\rho_0}\right)^{1/3}c^{4/3}t^{-4/3}\cr
&= 2.4\times 10^{42}R_{13}^2\kappa_{0.4}^{-4/3}\vt_{0,9}^{1/3}\rho_{0,-9}^{-1/3}t_{\rm hr}^{-4/3}\erg\se^{-1},
\end{align}
where $\vt_{0,9}=10^9\vt_0\cm\se^{-1}$, $\rho_{0,-9}=10^{-9}\rho_0 \gr\cm^{-3}$ and $t=1t_{\rm hr}$~hr. The weak dependence on the parameters $\rho_0$ and $\vt_0$ implies that, if detected, this power law tail can be used for an accurate determination of the stellar radius.

We emphasize that since $t_{\rm smear}\geq R/c$ and $t_{\rm sph}\sim R/4\vt_0$, the time interval  $t_{\rm smear}\ll t\ll t_{\rm sph}$ exists only for small velocities satisfying $\vt_0/c\ll1/4$. This condition is met only for large RSG progenitors. Given that $\vt_0/\rho_0\propto R^{1.38}$ for $n=3/2$, it is useful to rewrite the relation \eqref{eq:Lasymp} as
\begin{align}\label{eq:RLasymp}
R_{\rm late-L}&=\left(\frac{3L}{4\pi}\right)^{2/5}t^{8/15}(\kappa/c)^{8/15}\left(\frac{\rho_0R^{3/2}}{\vt_0}\right)^{2/15}\cr
&= 1.8\times 10^{13} L_{43}^{2/5}t_{\rm hr}^{8/15}\cm\cr
&~~\times\left(\frac{\rho_{0,-9}R_{13}^{3/2}}{\vt_{0,9}}\right)^{2/15}\kappa_{0.4}^{8/15},
\end{align}
where $L=10^{43}L_{43}\erg\se^{-1}$ and the factor in the last line is close to unity for RSG parameters. Using Eqs \eqref{eq:MVParam_v} and \eqref{eq:MVParam_rho},
\begin{equation}
\left(\frac{\rho_{0,-9}R_{13}^{3/2}}{\vt_{0,9}}\right)^{2/15}= 1.1 M_{10}^{0.026}\vt_{*,8.5}^{-0.24}R_{13}^{0.015}\kappa_{0.4}^{0.43} f_{\rho}^{0.07}~~(RSG).
\end{equation}

\section{Light curve}\label{sec:lightcurve}
In this section, the observed bolometric light curve is calculated assuming that the breakout is strictly spherically symmetric (i.e. assuming negligible difference in shock arrival times to the surface at different locations). In \sref{sec:LightTravelTime} exact expressions for the observed luminosity are given. In \sref{sec:LightCurveCharacteristics}, an approximate analytic expression (Eq. \eqref{eq:LobsAn2}) is derived assuming $ct_0/R\ll1$, which is valid for all progenitors with the exception of the largest RSGs. Approximate expressions for the peak luminosity and the luminosity at the time $R/c$ are given. The results of this section are summarized in \sref{sec:LightTravelSummary}.

\subsection{Exact light curve}\label{sec:LightTravelTime}
Due to the difference in arrival times of photons originating from different positions on the surface of the star, the actual luminosity $\Lobs(t)$, that a distant observer would measure, is related to the instantaneous luminosity $L(t)$ by
\begin{equation}\label{eq:LightTravelTimemu}
\Lobs(t)=\int_0^{1} h(\mu)L(t-R(1-\mu)/c)\mu d\mu,
\end{equation}
where $h(\mu)=2\pi I(\mu)|_{\tau=0}/\mathcal{L}$ is the angular distribution of the radiation intensity at the surface normalized so that
$\int_0^1 h(\mu)\mu d\mu=1$.

The precise value of $h(\mu)$ requires the solution of radiation transport up to the surface. For the non-relativistic breakouts considered here, the transport equations can be solved in the steady state approximation with the flux given by the diffusion solution. For the considered case of Thomson scattering, $h(\mu)$ was obtained analytically by  \citet[][]{Chandrasekhar50} and can be fit by a linear relation,
\begin{equation}\label{eq:LinearIntensity}
h(\mu)\approx a_{I}+b_{I}\mu, \quad a_I/2+b_I/3=1,
\end{equation}
with
\be a_I=0.85,\quad b_I=1.725,\label{eq:Chandra}\ee  to a good approximation (better than $3\%$, see \sref{sec:Chandra}).
For comparison, Eq. \eqref{eq:LinearIntensity} with $a_I=2,~b_I=0$ represents a black body surface (isotropic emission), while $a_I=1,~b_I=3/2$ represents isotropic scattering in the Eddington approximation. The resulting light curves for a density power law index $n=3$ and for different values of $ct_0/R$ are plotted in figures \ref{fig:L_obs_vs_s} and \ref{fig:L_obs_vs_t}.

\subsection{Approximate light curve for $R/c\gg t_0$}\label{sec:LightCurveCharacteristics}
For most progenitors, the smearing time scale $R/c$ is much larger than $t_0$.
In fact, using Eqs. \eqref{eq:MVParam_v} and \eqref{eq:MVParam_rho} we have
\begin{align}\label{eq:ct0R}
ct_0/R&= 0.02 M_{10}^{-0.45}\vt_{*,8.5}^{-1.45}R_{12}^{0.9}\kappa_{0.4}^{-0.45} f_{\rho}^{-0.18}~~(BSG)\cr
&= 0.05 M_{10}^{-0.58}\vt_{*,8.5}^{-1.58}R_{13}^{1.16}\kappa_{0.4}^{-0.58} f_{\rho}^{-0.28}~~(RSG).
\end{align}
As can be seen, except for very large RSG's with $R\sim 10^{14}\cm$, $ct_0/R$ may be assumed to be small.

In this case, the burst time scale is $R/c$ and the typical luminosity is $E_{\infty}c/R$.
It is useful to describe $\Lobs$ as a function of
\begin{equation}\label{eq:sdef}
s\equiv\frac{c(t-\tpeakj)}{R}.
\end{equation}
Consider first the formal limit $t_0c/R\ra0$. In this limit $L(t)\ra E_{\infty}\delta(t-\tpeakj)$ and the observed luminosity \eqref{eq:LightTravelTimemu} goes to
\begin{equation}\label{eq:t_0ra0}
\Lobs(t)\xra[t_0c/R\ra0]{}\frac{E_{\infty} c}{R}\left(1-s\right)h(1-s)\cdot (0<s<1).
\end{equation}

\begin{figure}[h]
\epsscale{1} \plotone{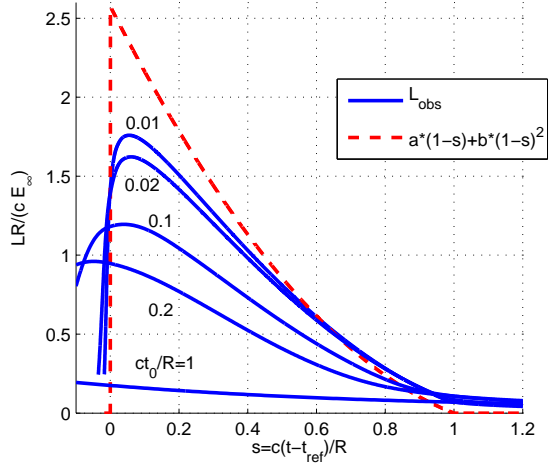}
\caption{Normalized observed luminosities as a function of the parameter $s=c(t-t_{\rm ref})/R$ for different stellar radii (blue solid lines), compared to the limit of Eq. \eqref{eq:t_0ra0} (dashed red). \label{fig:L_obs_vs_s}}
\end{figure}

\begin{figure}[h]
\epsscale{1} \plotone{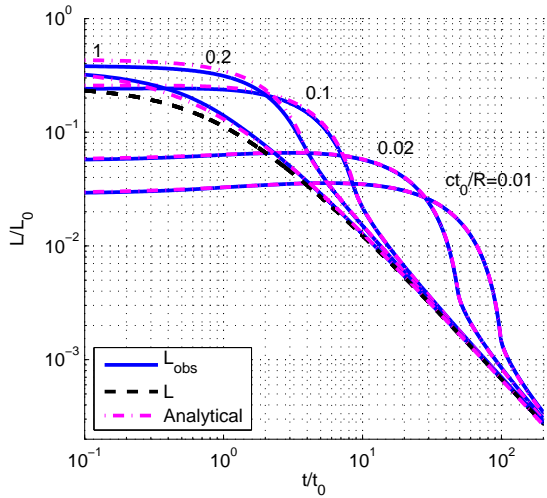}
\caption{Normalized observed luminosities (solid blue lines) as a function of time since (expected) breakout, for different star radii. Also plotted are the instantaneous emitted luminosity (black dashed line) and the analytical observed luminosity given by Eqs. \eqref{eq:LobsAn2} and \eqref{eq:gs} (magenta dashed-dotted line). \label{fig:L_obs_vs_t}}
\end{figure}

In figure \ref{fig:L_obs_vs_s}, the light curves are shown as a function of $s$ (blue solid lines) while the limit of Eq. \eqref{eq:t_0ra0} is shown for comparison (dashed red).  As can be seen, the observed luminosity converges slowly with $ct_0/R$ to the limiting value. This is due to the luminosity tale of $L\propto t^{-4/3}$ at late times, the integral of which converges slowly. This can be taken into account by using the approximation of Eq. \eqref{eq:ELApprox}. Using Eqs. \eqref{eq:LinearIntensity}, \eqref{eq:LightTravelTimemu} and \eqref{eq:ELApprox} we find
\begin{equation}\label{eq:LobsAn2}
\Lobs(t)=\frac{cE_{\infty}}{R}s_c^{1/3}\inv3\int_{0}^{\min(1,s-s_c)}(1-s')[a_I+b_I(1-s')](s-s')^{-4/3}ds',
\end{equation}
where
\be s_c=ca_tt_0/R.\label{eq:scdef}\ee An explicit expression for the integral in \eqref{eq:LobsAn2} is given in Eq. \eqref{eq:gs} and the resulting observed luminosities are shown in figure \ref{fig:L_obs_vs_t} (dashed dotted magenta lines, values of $E_{\infty}/E_0=2.03$ and $a_t=0.1$ fitted for the case $n=3$ considered were used). As can be seen in the figure, this is an excellent approximation to the calculated observed luminosity.

A few properties of the light curve can be derived from Eq. \eqref{eq:LobsAn2}. At very large $s\gg1$ Eq. \eqref{eq:LobsAn2} reduces to equation \eqref{eq:Lasymp} as required.
Given that $s_c\ll1$, the value of $\Lobs$ at $s=1$ ($t=\tpeakj+R/c$) is
\begin{equation}\label{eq:LobsRoC}
\Lobs(t=\tpeakj+R/c)=\frac{cE_{\infty}}{R}s_c^{1/3}(a_I/2+b_I/5),
\end{equation}
where for $a_I=0.85(2)$ we have $(a_I/2+b_I/5)=0.77(1)$.
Comparing to \eqref{eq:Lasymp} (and using the relation between $a_t,E_{\infty},L_{\infty}$ in \eqref{eq:Einfty}), we see that at $t=R/c$ the luminosity drops to a value about $3$ times larger than the extrapolation of the asymptotic luminosity (Eq. \eqref{eq:Lasymp}) to this time.

As shown in \sref{sec:gs}, the peak observed luminosity is, to a good approximation, given by (Eq. \eqref{eq:LobsmaxAnalytic})
\begin{equation}\label{eq:Lpeak}
L_{\rm Peak}=(a_I+b_I)\frac{E_{\infty}c}{R}\left[1-\left(\frac{ct_0}{R}\right)^{1/4}\right].
\end{equation}

\subsection{Light curve calculation summary}\label{sec:LightTravelSummary}
The observed light curve can be calculated using Eqs \eqref{eq:LightTravelTimemu} and \eqref{eq:LinearIntensity}, with $L(t)$ given by \eqref{eq:LNorm} and tabulated in \citet[][, table 3]{Sapir11}. The following parameters completely determine the light curve:  breakout luminosity $L_0$, breakout shock crossing time $t_0$, time of peak planar flux $\tpeakj$, progenitor radius $R$ and density power law index $n$. As shown in \citet{Sapir11}, the luminosity depends weakly on $n$. The parameters $L_0$ and $t_0$ are related to the breakout density, velocity and opacity through Eqs. \eqref{eq:L_0def} and \eqref{eq:tNorm}.
The resulting light curves for the case $n=3$ are shown in figure \ref{fig:L_obs_vs_t}.

For most progenitors $R/c\gg t_0$ and the observed light curve can be calculated using Eq. \eqref{eq:LobsAn2} with $s$ given by Eq.~\eqref{eq:sdef} and $a_l=0.85,~b_l=1.725$ given by Eq.~\eqref{eq:Chandra}. An explicit algebraic expression is given in Eq. \eqref{eq:gs}. In this case the light curve is determined by the following parameters: breakout energy $E_{\infty}$, time of peak planar flux $\tpeakj$, progenitor radius $R$ and a dimensionless parameter $s_c$. The parameters $E_{\infty}$ and $s_c$ are related to the breakout density, velocity and opacity (and $R$) through Eqs. \eqref{eq:scdef}, \eqref{eq:tNorm} and \eqref{eq:Einfty}.

The peak luminosity in this case can be approximated by Eq. \eqref{eq:Lpeak}. At $t=\tpeakj+R/c$ the luminosity drops by a factor $\sim(0.1ct_0/R)^{1/3}$ compared to the peak, to a value which is approximately 3 times higher than the extrapolation (to $t=\tpeakj+R/c$) of the asymptotic luminosity given by Eq.~\eqref{eq:Lasymp}.

In the extreme limit $t_0\ra0$ a simple approximation for the light curve (for $t<R/c$) is given by \eqref{eq:t_0ra0}, which depends on two parameters only, $E_{\infty}$ and $R$.

\section{Comparison to previous studies} \label{sec:Comparison} 
In \sref{sec:Ensman92} we compare our results with those of the numerical calculations of \citet{Ensman92} for 1987A like (BSG) progenitors. In \sref{sec:AnalyticCompare} we compare our calibrated analytic results with the order of magnitude estimates of \citet{Matzner99} and \citet{Nakar10}.

\subsection{Comparison to \citet{Ensman92}}\label{sec:Ensman92}
\begin{figure}[h]
\epsscale{1} \plotone{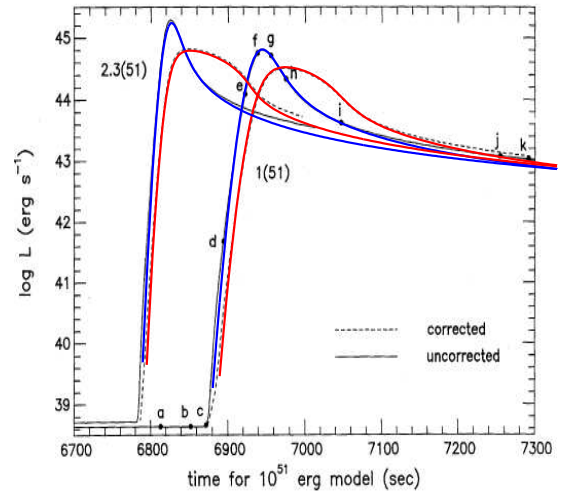}
\caption{A comparison of our light curves with those obtained by the numerical calculations of \citet{Ensman92} for two progenitor models ('500full1' and '500full2'). Our results are shown in color (blue and red for source frame and observed luminosities respectively), overlaid on the original figure~2 of \citet{Ensman92}. For the comparison we have used  $e+p=2.5\mathcal{L}/c$ and $h(\mu)=2$, as used by \citet{Ensman92}, although these values do not yield an accurate description of the light curves (see text). \label{fig:L_vs_tEB}}
\end{figure}

In figure~\ref{fig:L_vs_tEB} we compare our results with the numerical light curves of \citet{Ensman92}. As can be seen, the numerical light curves are in excellent agreement with our analytic light curves, for an appropriate choice of $\rho_0$ and $\vt_0$. The fitted values are $\{\vt_0=16,500\km\se^{-1},~~ \rho_0=1.6\times 10^{-9}\}$ and $\{\vt_0=25,000\km\se^{-1},~~\rho_0=1.2\times10^{-9}\}$ for the progenitor models '500full1' and '500full2' respectively (the opacity is assumed to be $\kappa=0.33\cm^{2}\gr^{-1}$, appropriate to a mixture of ionized Hydrogen and Helium with mass fractions $X=0.67$ and $Y=0.33$).

\citet{Ensman92} have used $e+p=2.5\mathcal{L}/c$ for the calculation of the instantaneous lab frame luminosity, $L_{\rm lab}=4\pi R^2[\mathcal{L}+\vt(e+p)]$, which is not accurate to first order in $\bt$ (see \sref{sec:FirstOrder} for a careful calculation of the frame transformation), and assumed isotropic emission ($h(\mu)=2$) for the light travel time smearing calculation, which does not represent Thompson scattering opacity (see \S~\ref{sec:LightTravelTime}). For the comparison we have adopted similar values in our calculation of the light curves shown in figure~\ref{fig:L_vs_tEB}, although they do not yield an accurate description of the light curves.

Using Eq. \eqref{eq:t_sph}, the transition to the spherical phase is expected at approximately  $7400\se$ and $7100\se$ respectively \citep[using the reference time as in figure 2 in][]{Ensman92}. As can be seen in figure \ref{fig:L_vs_tEB}, the planar solution agrees with the spherical calculation at the latest times presented (which are less than $t_{\rm sph}$) up to a factor of $\sim1.5$. The maximal velocities predicted by equation Eq. \eqref{eq:vmax} are $28,000\km\se^{-1}$  and $43,000\km\se^{-1}$, in good agreement (up to $\sim 10\%$) with the maximal velocities of $30,000\km\se^{-1}$ and $48,000\km\se^{-1}$ obtained in \citet{Ensman92}.

\subsection{Comparison with previous analytical work}\label{sec:AnalyticCompare}
A quantitative comparison of the analytic results presented here \citep[calibrated using the numerical results of][]{Sapir11} to those of the order of magnitude estimates made in the literature is complicated since the breakout parameters $\rho_0$ and $\vt_0$ were not well defined in earlier analyses. We remind the reader that we define $\rho_0$ and $\vt_0$ to be the upstream density and the shock velocity that would be obtained in a pure hydrodynamic calculation (ignoring radiation diffusion) at the point where $\tau=c/\vt_s$ (see \sref{sec:Planar}). For illustration purposes we compare a few of our results with order of magnitude estimates made by \citep{Matzner99} and \citep{Nakar10} in which $\rho_0$ and $\vt_0$ were (vaguely) defined in a similar way.
\begin{itemize}
\item The numerical values of the order of magnitude estimates of the power law decline of the luminosity, $L\propto t^{-4/3}$, made by \citep{Nakar10} can be summarized as $L(t)\sim 4\pi R^2\kappa^{-4/3}(\vt_0/\rho_0)^{1/3}c^{4/3}t^{-4/3}$ (using the values for $\rho_0$ and $\vt_0$ given in their appendix). This result is in rough agreement with our Eq. \eqref{eq:Lasymp}, with their expressions overestimating the luminosity by a factor of about $3$.
\item \citet{Matzner99} estimate that the maximal velocity obtained by the fastest ejecta is about twice the breakout shock velocity  $\vt_0$. Coincidently (as far as we can tell), this turns out to be accurate (see \sref{eq:vmax}).
\item The numerical values for the breakout energy of RSGs and BSGs in \citep{Matzner99} are related to the numerical values of the maximal velocities by $E_{\rm BO}=4\pi R^2 c\vt_{\max}/\kappa$ and $E_{\rm BO}=1.7\times 4\pi R^2 c\vt_{\max}/\kappa$ respectively. The RSG relation is accurate while in the BSG relation the emitted energy is overestimated by a (modest) factor of about $1.7$ (see \eqref{eq:vmaxEBr}).
\end{itemize}

\section{Summary \& discussion}\label{sec:Summary}

We have derived exact bolometric light curves of supernova shock breakouts using the universal planar breakout solutions \citep{Sapir11}, assuming spherical symmetry, constant Thomson scattering opacity, $\kappa=\ave{Z/A}\sig_T/m_p$, and angular intensity corresponding to the constant flux limit. The light curves are insensitive to the form of the density profile. This was demonstrated by calculating the emission for power law profiles $\rho\propto x^{n}$ with resulting luminosities changing by $<30\%$ for a broad range of power law indexes, $1<n<10$ \citep[][]{Sapir11}. 

The breakout emission properties are determined by four dimensional parameters: the progenitor radius $R$, the breakout velocity and density, $\vt_0$ and $\rho_0$ respectively, and $\kappa$. $\vt_0$ and $\rho_0$ are the shock velocity and (pre-shock) density at the point where $\tau_{\rm sh}=c/\vt_{\rm sh}$ is reached in the pure hydrodynamic (neglecting radiation diffusion) solution (see \S~\ref{sec:Planar} for exact definitions). The relations between the SN parameters, the ejecta mass $M_{\rm ej}$ and bulk velocity $\vt_*$, and the breakout parameters, $\rho_0$ and $\vt_0$, are given in Eqs.~\eqref{eq:MVParam_v} and~\eqref{eq:MVParam_rho} \citep[see also][]{Matzner99,Katz09,Nakar10}.

The application of the planar solution to SN breakouts was discussed in \sref{sec:Planar_SNe}. The planar approximation is applicable provided that the shock crossing time at breakout, $t_0=c/\kappa\rho_0\vt_0^2$, is much smaller than the time for transition to spherical expansion, $t_{\rm sph}\sim R/(4\vt_0)$ \citep[Eq. \eqref{eq:t_sph},][]{Piro10,Nakar10}. This is valid for practically all progenitors, see Eq.~(\ref{eq:t_0t_sph}). At $t>t_{\rm sph}$ the expansion is no longer planar, and the planar approximation no longer holds. Our results are not applicable for progenitors with optically thick winds.

Deviation from spherical symmetry may affect breakout light curves due to several effects: the breakout velocity may be different at different surface locations, the shock my reach the surface at oblique angles, and the shock arrival time at the surface may depend on location. In this paper we have assumed that the shock reaches the surface with approximately the same velocity everywhere and parallel to the surface. Differences in shock arrival times to the surface, $\Delta t_{\rm asym}$, due to moderately asymmetric explosions, affect the early light curve \citep[e.g.][]{Calzavara04}, at $t<\Delta t_{\rm asym}$, but do not affect the maximal velocity of the ejecta and the total emitted energy, $\vt_{\max}$ and $E_{\rm BO}$.

Analytic expressions for $\vt_{\max}$, $E_{\rm BO}$ and the late time luminosity were derived in section \sref{sec:EnergyVelocityLuminosity}. The total energy of the breakout burst is approximately (Eq. \eqref{eq:EBreakout})
\be E_{\rm BO}= 8\pi R^2\kappa^{-1}\vt_0 c=1.9\times 10^{47}\erg R_{13}^2\vt_9\kappa_{0.4}^{-1}.
\end{equation}
where $\vt_0=10^9\vt_9\cm\sec^{-1}$. 
We have shown that integrating the luminosity to times greater than $t_{\rm sph}$ affects the total energy considerably only at very late times, $t\gtrsim R/\vt_{*}$ (Eq. \eqref{eq:DtBr}). The maximal velocity of the ejecta is directly related to the emitted energy by \citep[Eq. \eqref{eq:vmaxEBr}, see also][]{Lasher79,Sapir11}
\be\vt_{\max}= \kappa E_{\rm BO}/(4\pi R^2c)= 2.0\vt_0\,.\ee 
We note that these results are not sensitive to deviations from of the steady state flux angular intensity distribution we have used. For $4\vt_0\ll c$, valid for large RSG progenitors, there is a significant separation between $R/c$ and $t_{\rm sph}$. In this case, the luminosity at $\max(\Delta t_{\rm asym},R/c)< t<t_{\rm sph}$ is approximately given by (Eq. \eqref{eq:Lasymp})
\be L(t)= (4/3)\pi R^2(\vt_0/\rho_0)^{1/3}(c/\kappa)^{4/3}t^{-4/3}.\ee 
The strong dependence of the asymptotic luminosity on $R$ and weak dependence on $\rho_0$ and $\vt_0$ allows one to accurately determine the progenitor radius of RSG breakouts (Eq. \eqref{eq:RLasymp}), 
\be R= 2\times10^{13}L_{43}^{2/5}t_{\rm hr}^{8/15}\cm.\ee

The bolometric light curve, assuming negligible spread in shock arrival times, are calculated in \sref{sec:lightcurve}. A proper calculation of the effects of finite light travel time requires knowledge of the angular dependence of the intensity. Fortunately, the problem of radiation transport in an optically thick medium with opacity dominated by Thompson scattering was solved in closed form \citep[][results summarized in \sref{sec:Chandra}]{Chandrasekhar50}. Exact light curves can be calculated using Eqs.~\eqref{eq:LightTravelTimemu} and~\eqref{eq:LinearIntensity} and the planar luminosity functions, $\mathcal{L}$, given in \citep{Sapir11}. Some examples are shown in figures \ref{fig:L_obs_vs_s} and \ref{fig:L_obs_vs_t}. For cases where $ct_0\ll R$, applicable in all progenitors except for the largest RSGs (see Eq.~\ref{eq:ct0R}), the planar luminosity  $\mathcal{L}$ can be approximated by a power law, Eq.~\eqref{eq:ELApprox}, allowing the derivation of an analytical expression for the light curve, given in Eqs.~\eqref{eq:LobsAn2} and~\eqref{eq:gs}. The analytic expression is compared to the exact calculation (both without relativistic corrections) in figures~\ref{fig:L_obs_vs_s} and~\ref{fig:L_obs_vs_t}. In this case, the peak luminosity is approximately given by
\be L_{\rm obs, peak} = 2.5(E_{\rm BO}c/R)[1-(ct_0/R)^{1/4}],\ee
where
\be  E_{\rm BO}c/R=5.6\times 10^{44}R_{13}\vt_9\kappa_{0.4}^{-1}\erg\sec^{-1}\ee
is the typical peak luminosity. 
In addition, we have shown that at $t=R/c$ the normalized luminosity $Lt^{4/3}$ is about $3$ times larger than its asymptotic value given by Eq. \eqref{eq:Lasymp}. A short summary which explains how to use the different expressions to obtain the light curves is provided in \sref{sec:LightTravelSummary}.

The transformation of the rest frame intensity to the lab frame and the value of the retarded time introduces corrections of order $\bt^1$. These are calculated in \sref{sec:FirstOrder}. For the velocities considered, $\bt_0\lesssim0.3$, the only considerable correction is to the early light curve and is given by \eqref{eq:FirstOrderLc}. We note that other corrections of order $\bt^1$ are not excluded. 

The results of this paper are compared to previous results in \sref{sec:Comparison}. The calculated bolometric light curve is shown to be in excellent agreement with the numerical calculation of \citet{Ensman92}, see figure~\ref{fig:L_vs_tEB}. The order of magnitude estimates given by \citet{Matzner99} and by \citet{Nakar10} for the emitted energy, maximal velocity and asymptotic luminosity agree to within factors of few with our exact analytic expressions.

The properties of the breakout emission depend strongly on the radius of the progenitor $R$ and on the breakout shock velocity $\vt_0$, depend weakly on the value of the density at breakout $\rho_0$, and are insensitive to the density structure. Breakout observations therefore allow one to accurately determine $R$ and $\vt_0$.  These quantities are directly related to other observables: The breakout shock velocity $\vt_0$ is roughly proportional to the ejecta velocity $\vt_*$, which is probed by SN observations. The maximal velocity of the ejecta (about twice the breakout shock velocity) can be constrained by the radio and X-ray emission produced by the interaction of the ejecta with the circumstellar medium \citep[e.g.][]{Waxman07,Soderberg08}. Both parameters affect the subsequent spherical expansion (cooling envelope) phase of the emission \citep[e.g.][]{Chevalier92,Rabinak10,Nakar10}.

\acknowledgements We thank Adam Burrows and Subo Dong for useful discussions. This research was partially supported by Minerva, ISF, and the Universities Planning \& Budgeting Committee grants. B.K is supported by NASA through Einstein Postdoctoral Fellowship awarded by the Chandra X-ray Center, which is operated by the Smithsonian Astrophysical Observatory for NASA under contract NAS8-03060.

\appendix
\section{A. Light travel time averaging}\label{sec:gs}
The integral
\begin{equation}
g(s;s_c)=s_c^{1/3}\inv3\int_{0}^{\min(1,s-s_c)}(1-s')[a_I+b_I(1-s')](s-s')^{-4/3}ds' 
\end{equation}
appearing in equation \eqref{eq:LobsAn2} is explicitly given by
\begin{equation}\label{eq:gs}
g(s;s_c)=\left\{\begin{array}{cc}
\frac{3}{2}s_c^{1/3}\left[- a_I (s-1)^{2/3} + \frac65 b_I (s-1)^{5/3}- \left[a_I (\frac23 - s) + 2 b_I (\frac13 - s + \frac35 s^2)\right] s^{-1/3}\right]~~~~&s>1+s_c\\
&\\
a_I (1 - s - \inv2s_c) +  b_I \left[1  + (s_c-2)s + s^2 - s_c (1 + \inv5s_c)\right]+(s_c/s)^{1/3}\left[-a_I -b_I + \frac32 (a_I + 2 b_I) s - \frac{9}{5} b_I s^2\right]~~~~&s_c<s\leq1+s_c\\
&\\
0~~~~&s\leq s_c\\ \end{array}\right.
 \end{equation}

To lowest order in $s_c$,  the peak of $g(s;s_c)$ is reached at
\begin{equation}
s_{\rm peak}=\left(\frac{a_I+b_I}{3(a_I+2b_I)}\right)^{3/4}s_c^{1/4},
 \end{equation}
 with a peak value of
\begin{equation}
g_{\rm peak}=(a_I+b_I)\left[1-(3^{-3/4}+3^{1/4})\left(\frac{a_I+2b_I}{a_I+b_I}\right)^{1/4}s_c^{1/4}\right].
 \end{equation}
Using Eq. \eqref{eq:scdef}, the fact that for $0<a_I<2$ we have $1<[(a_I+2b_I)/(a_I+b_I)]^{1/4}<1.2$, and the numerical value  $(3^{-3/4}+3^{1/4})a_t^{1/4}=1.0$ (using $a_t=0.125$, see Eq. \eqref{eq:Einfty}), we conclude that to a good approximation
\begin{equation}\label{eq:LobsmaxAnalytic}
L_{\rm obs, \rm peak}=\frac{cE_{\infty}}{R}(a_I+b_I)\left[1-\left(\frac{ct_0}{R}\right)^{1/4}\right].
\end{equation}

\section{B. First order corrections due to the transformation between the comoving frame and the observer frame}\label{sec:FirstOrder}
In this section the first order corrections in $\bt$ due to the transformation between the comoving frame and the observer frame (lab frame) are calculated. Note that the expression for the asymptotic luminosity \eqref{eq:Lasymp} is valid only for small velocities $\vt_0\ll c/4$, for which the first order corrections are negligible. We estimate the first order corrections in $\bt_0=\vt_0/c$ to the value of the breakout energy, Eq. \eqref{eq:EBreakout}, in \S~\ref{sec:1stEBO}, and the corrections to the light curve in \S~\ref{sec:1st_lc}.

\subsection{B1. First order corrections to the Breakout Energy}\label{sec:1stEBO}
The rate of energy escaping from the surface of the expanding envelope as measured in the laboratory frame, $dE_{\rm lab}/dt$,  is given by
\begin{equation}\label{eq:dElabdt}
\frac{dE_{\rm lab}}{dt}=4\pi r^2(\mathcal{L}_{\rm lab}-\vt e_{\rm lab}),
\end{equation}
where $r(t)$ is the radius of the surface and $\mathcal{L}_{\rm lab}$ and $e_{\rm lab}$ are the radiation flux and energy density at the surface in the laboratory frame.  The values of $\mathcal{L}_{\rm lab}$ and $e_{\rm lab}$  can be expressed using the comoving frame flux $\mathcal{L}$,  energy density $e$ and pressure $p$ by

\begin{align}
&\mathcal{L}_{\rm lab}=\mathcal{L}+\vt (p+e)+O(\bt^2),\cr
&e_{\rm lab}=e+2\bt c^{-1} \mathcal{L}+O(\bt^2).
\end{align}
The emitted energy per unit time to first order in $\bt$  is thus
\begin{equation}\label{eq:dElabdt2}
\frac{dE_{\rm lab}}{dt}=4\pi r^2\left(1+\bt\left.\frac{pc}{\mathcal{L}}\right |_{\tau=0}\right)\mathcal{L}.
\end{equation}
The term $(pc/\mathcal{L})_{\tau=0}$ depends on the angular distribution of the radiation intensity at the surface, which depends on the transport properties of the medium. For Thomson scattering in the constant flux limit, the value is $(pc/\mathcal{L})_{\tau=0}=0.71$ \citep[][see also \sref{sec:Chandra}]{Chandrasekhar50}. For comparison, for a black body surface (isotropic emission), the value is $(pc/\mathcal{L})_{\tau=0}=2/3$ and for isotropic scattering in the Eddington approximation it is $(pc/\mathcal{L})_{\tau=0}=17/24$.

Assuming that $(pc/\mathcal{L})_{\tau=0}$ is constant over time and neglecting the change in the surface radius during the emission, Eq. \eqref{eq:dElabdt2} can be analytically integrated over time. To do this note that $4\pi R^2\int^{t} \mathcal{L}(t')dt'=E(t)$ and that
\begin{equation}\label{eq:IntLbt}
4\pi R^2\int^t \mathcal{L}(t')\vt(t')dt' = \frac{c}{\kappa}\int^t \frac{dE}{dt'}Edt'=\haf E(t)\vt(t),
\end{equation}
where Eq. \eqref{eq:v_to_Egen} was used. Using Eqs. \eqref{eq:IntLbt} the integration of equation \eqref{eq:dElabdt2} to infinity results in
\begin{equation}\label{eq:Elab}
E_{\rm lab,\infty}=E_{\infty}\left(1+ \haf\left.\frac{pc}{\mathcal{L}}\right |_{\tau=0}\bt_{\infty}\right),
\end{equation}
where $\bt_{\infty}=\vt_{\infty}/c$ is the asymptotic value of $\bt$ in the planar approximation (approximately unchanged by spherical geometry, see
\sref{sec:AssymVel}).
Using the approximation of Eq.~\eqref{eq:IntLbt}, and adopting $(pc/\mathcal{L})_{\tau=0}=0.7$, we have
\begin{equation}
E_{\rm lab,\infty}=E_{\infty}(1+ 0.7\bt_0)= E_{\infty}(1+ 0.35\bt_{\infty}).
\end{equation}
Linearly adding this correction and the small term $(t_{\rm sph}/(a_t t_0))^{-1/3}$ in equation \eqref{eq:EBreakout} we obtain
\begin{equation}\label{eq:EBreakout1order}
E_{\rm BO}=E_{\infty}\left[1-\left(\frac{t_{\rm sph}}{a_t t_0}\right)^{-1/3}+ 0.7\bt_0\right].
\end{equation}

\subsection{B2. First order correction to the observed Light curve}\label{sec:1st_lc}
First order corrections to the light curve arise from corrections to the intensity, direction and value of the retarded time. At times $t\gtrsim R/c$, the surface of the star moved a distance of $(\vt/c)R$ implying that there are corrections of order $\vt/c$ due to the spherical nature of the expansion. An estimate of this correction is beyond the scope of this paper. Here we focus on the first order corrections at early times $t\ll R/c$ (including the peak observed luminosity). At these early times the angle between the emitting region radius and the direction towards the observer is small, $\mu-1\ll1$, allowing a simple derivation of the first order correction.

Consider a spherically symmetric moving surface with radius $r(t)$ emitting radiation with lab frame intensity $I(t;\bOm)$ which is axisymmetric with respect to the surface normal. Consider an observer located in the direction $\bOm_{\rm obs}$ at a very large distance. The position on the sphere is parametrized by the angle $\theta$ between $\bOm$ and $\bOm_{\rm obs}$. Since the intensity $I$ is assumed to be the same at any position on the sphere and to be axisymmetric with respect to the normal we have $I(t,\bOm_{\rm obs})=I(t,\te)$.
 The luminosity inferred by the observer is given by
\begin{equation}\label{eq:Lobsda}
\Lobs(\tobs)=4\pi\int 2\pi a da I(t;\te),
\end{equation}
where
\begin{equation}\label{eq:a_te_vp}
a=r\sin\te
\end{equation}
and the lab (retarded) time $t$ is related to the observer time $\tobs$ through
\begin{equation}\label{eq:tobs_t}
R+ct-r\cos\te=c\tobs.
\end{equation}
The integral \eqref{eq:Lobsda} should be evaluated at constant $\tobs$, with $t$ and $\te$ functions of $a$ through the relations \eqref{eq:a_te_vp} and  \eqref{eq:tobs_t}, in the regime $\bt<\cos\te<1$ (it is assumed that there are no photons with $\cos\te<\bt$, which are coming from outside of the surface).

It is useful to solve for $\mu=\cos\te$
\begin{equation}\label{eq:muoftpr}
\mu=\frac{R-c(\tobs-t)}{r},
\end{equation}
where $t'=\tobs-t$. Using \eqref{eq:tobs_t}, we find
\begin{equation}
cdt=rd\mu+\mu\vt dt\soo rd\mu= c(1-\bt\mu)dt
\end{equation}
and
\begin{align}
ada&=r\sin^2\te\vt dt+r^2\sin\te\cos\te d\te\cr
&=cr[(1-\mu^2)\bt-\mu(1-\bt\mu)]dt=cr(\bt-\mu)dt.
\end{align}
Equation \eqref{eq:Lobsda} can be written as
\begin{equation}\label{eq:Lobsdt}
\Lobs(\tobs)=4\pi\int dt cr(\mu-\bt)2\pi I(t;\mu)
\end{equation}
with $\mu$ given by \eqref{eq:muoftpr}.

The lab frame $\mu$ is related to the surface frame $\mu$, $\mucom$, by
\begin{equation}\label{eq:mucom}
\mucom=\frac{\mu-\bt}{1-\bt\mu},
\end{equation}
while the lab frame intensity $I$ is related to the surface frame $\Icom$ by
\begin{equation}
I(t,\mu)=[\gamma(1+\bt\mucom)]^3\Icom(t,\mucom),
\end{equation}

where $\gamma=(1-\bt^{2})^{1/2}$ is the surface Lorenz factor.

Assuming that the angular dependence of the comoving intensity is time independent, $\Icom(t,\mu)=h(\mu)j_{e,\rm com}(t)/(2\pi)$, equation \eqref{eq:Lobsdt} can be written as
\begin{equation}
\Lobs(\tobs)=\int dt  \frac{c}{r}(\mu-\bt)\gamma^3(1+\bt\mucom)^3 h(\mucom)\Lcom(t),
\end{equation}
where $\mu$ and $\mucom$ are given in eqs \eqref{eq:muoftpr} and \eqref{eq:mucom}.

Consider next the first order corrections in the parameters $\bt$ and $(1-\mu)$, neglecting the difference between $r$ and $R$. Note that in this approximation we can set $\bt\mu\ra\bt,~\gamma\ra 1, ~\mucom\ra\mu$ and
\begin{equation}\label{eq:fsmearCorr}
\Lobs(\tobs)= \frac{c}{R}\int dt \mu h(\mu)(1+2\bt)\Lcom
\end{equation}
with $\mu=1-c(\tobs-t)/R$.

Using $\int_{-\infty}^t L(t)\vt(t)dt\approx \haf\vt_{\infty}E_{\infty}\approx 2\vt_0E_\infty$ (see \eqref{eq:IntLbt}), we find that in the limit $t_0c/R\ll1$, the correction at early times $\tobs\ll R/c$ is
\be\label{eq:FirstOrderLc}\Lobs=L_{\rm obs,0}(1+2\bt_0).\ee

\section{C. Steady state radiation transport with Thomson scattering}\label{sec:Chandra}
The steady state problem of radiation transfer in a semi-infinite medium with Thompson scattering was analytically solved by \citep{Chandrasekhar50}. Assuming a constant flux $\mathcal{L}$ coming from inside the medium and zero incident radiation, the intensities at the surface, $I_l(\mu)$ and $I_r(\mu)$, polarized in and perpendicular to the meridian plane respectively,  were found in closed form. The polarized intensities are given by
\begin{align}
I_l&=\frac{q}{\sqrt{2}}\frac{3}{8\pi}\mathcal{L}H_l(\mu),\cr
I_r&=\frac{1}{\sqrt{2}}\frac{3}{8\pi}\mathcal{L}H_{r}(\mu)(\mu+c),\cr
\end{align}
where $H_i$ ($i=l,r$) are the solutions to the integral equations
\begin{equation}
H_i(\mu)=1+\mu H_i(\mu)\int_0^1\frac{a_i(1-\mu^2)}{\mu+\mu'}H_i(\mu')d\mu'
\end{equation}
with $a_l=3/4$ and $a_r=3/8$, while $q$ and $c$ are the solutions to the equations
\begin{align}
q^2&=2(1-c^2),\cr
qH_l(1)&=(1+c)H_r(1).\cr
\end{align}
The resulting intensities were numerically calculated by Chandrasekhar and his secretary, Mrs. Frances H. Breen, to  5 decimal points for the twenty values of $\mu=0:0.05:1$ using pen and paper \citep[Table XXIV in][note that $F=\mathcal{L}/\pi$]{Chandrasekhar50}. The resulting total intensity, $I=I_l+I_r$, can be fit by the linear relation
\begin{equation}\label{eq:ChandraThompsonLin}
h_{Ch}(\mu)=\frac{2\pi I}{\mathcal{L}}\approx 0.85+1.725\mu
\end{equation}
to an accuracy better than $3\%$ for all $0<\mu<1$ (by assumption, $I(\mu)=0$ for $\mu<0$) .

\bibliographystyle{apj}

\end{document}